\documentclass[letterpaper]{article}

\usepackage[utf8]{inputenc} 
\usepackage[T1]{fontenc}
\usepackage{geometry}
\usepackage{setspace}
\usepackage{gensymb} 

\usepackage[style = chem-acs, autocite=superscript, doi=false, url=false]{biblatex}

\bibliography{Mo2C-plainbib}

\usepackage{graphicx}
\usepackage{float}
\newfloat{scheme}{htbp}{los}
\floatname{scheme}{Scheme}
\floatname{chart}{Chart}
\newfloat{graph}{htbp}{loh}

\usepackage{chemformula} 
\usepackage[version = 4]{mhchem} 

\DeclareUnicodeCharacter{2212}{-}
\usepackage{authblk}
\author[1]{Wei Gao}
\author[1]{Kaixuan Fan}
\author[2,3]{Menghan Li}
\author[2,3]{Jinhao Cheng}
\author[1]{Peng Zhu}
\author[2,3]{Qing Zhang}
\author[2,3,4]{Shuaishuai Ding}
\author[2,3,6]{Wenping Hu}
\author[1]{Fan Yang}
\author[2,3,5]{Dechao Geng}
\author[1,6]{Hechen Ren*}
\affil[1]{Center for Joint Quantum Studies \& Tianjin Key Laboratory of Low Dimensional Materials Physics and Preparing Technology, Department of Physics, School of Science, Tianjin University, Tianjin 300072, China}
\affil[2]{State Key Laboratory of Advanced Materials for Intelligent Sensing, Ministry of Science and Technology \& Key Laboratory of Organic Integrated Circuit, Ministry of Education \& Tianjin Key Laboratory of Molecular Optoelectronic Sciences, Department of Chemistry, School of Science, Tianjin University, Tianjin 300072, China}
\affil[3]{Collaborative Innovation Center of Chemical Science and Engineering, Tianjin 300072, China}
\affil[4]{Institute of Molecular Aggregation Science, Tianjin University, Tianjin 300072, China}
\affil[5]{Haihe Laboratory of Sustainable Chemical Transformations, Tianjin 300192, China}
\affil[6]{Joint School of National University of Singapore and Tianjin University, International Campus of Tianjin University, Binhai New City, Fuzhou 350207, China}

\title{Observation of field-odd and field-free superconducting diode effects in $\mathrm{Mo}_2\mathrm{C}$ nanoflakes}
\date{*Email: ren@tju.edu.cn}

\begin{document}
\begin{spacing}{2} 

\maketitle
\clearpage
\begin{abstract}
The superconducting diode effect (SDE) enables nonreciprocal supercurrent flow, holding immense potential for ultra-low-power quantum electronics. Intrinsic SDE typically requires materials with inherent symmetry breakings. Here, we report the discovery of SDE in chemical vapor deposition-grown molybdenum carbide ($\mathrm{Mo}_2\mathrm{C}$) nanoflakes, a material traditionally considered centrosymmetric. Strikingly, this system uniquely hosts both field-odd and field-free SDEs. Transport measurements reveal a field-odd SDE with tunable efficiency exceeding 40\% at 4 K under a perpendicular in-plane magnetic field. In a separate sample, a robust field-free SDE persists under zero-field and field-coolings. Out-of-plane field sweeps confirm the intrinsic nature of these phenomena. We propose that domain-boundary supercurrents or charge density wave-like orders drive this unexpected combination of symmetry breakings. Our findings establish air-stable $\mathrm{Mo}_2\mathrm{C}$ as an ideal platform for nonreciprocal superconducting electronics operating at liquid-helium temperatures, expanding the search for SDE into nominally centrosymmetric superconductors.
\end{abstract}

\section*{Keywords}

superconducting diode effect, unconventional superconductivity, nonreciprocal transport, 2D materials, CVD, MXene.


\section{Introduction}
Superconducting diode effect (SDE), characterized by a non-reciprocal critical current, allows for lossless current flow in one direction while exhibiting resistance in the other \autocite{Ando2020, Nadeem2023, Moll2023, Ma2025}. Unlike semiconductor $p-n$ junctions, SDE operates within the superconducting regime, offering a pathway toward ultra-low-power logic and memory devices for quantum information processing \autocite{He2022a}. Especially interesting among the SDE discoveries are the intrinsic ones \autocite{Daido2022}, as opposed to the ubiquitous SDE, which can appear due to accidental edge asymmetry in conventional superconducting films \autocite{Hou2023}. Unlike its cousin, the Josephson diode effect, where the Josephson phase plays a leading role in the nonreciprocal transport \autocite{Davydova2022, Davydova2024}, intrinsic SDEs rely on inversion symmetry breaking within the bulk of the sample. Additionally, breaking time-reversal symmetry (TRS) often leads to finite-momentum Cooper pairs and could correlate with unconventional superconductivity and pair density waves (PDW) \autocite{Gu2023, Zhao2023b, Kong2025, Han2025a}.

Originally, intrinsic SDE was discovered in "artificial" materials such as $[\mathrm{Nb/V/Ta}]_n$ superlattices \autocite{Ando2020}, Nb/V/Co/V/Ta multilayers \autocite{Narita2022}, and ferromagnetic/superconducting heterostructures \autocite{Hu_2025}. Lately, SDE has also been observed in van der Waals (vdW) materials with broken inversion symmetry and strong spin-orbit coupling such as $\mathrm{NbSe}_2$ \autocite{Bauriedl2022}, twisted trilayer graphene \autocite{Lin2022}, high-$T_\mathrm{C}$ cuprate flakes \autocite{Qi2025}, and chiral molecule–intercalated $\mathrm{TaS}_2$ \autocite{Wan2024}. Usually, in an SDE, TRS is broken by an external field or magnetic/chiral orders within the material platform, with the latter case referred to as the field-free \autocite{Li2024f, Narita2022} or zero-field SDE \autocite{Lin2022, Hu_2025}. Under certain strained conditions, SDE can even occur without breaking TRS \autocite{Liu2024b}. However, all these materials feature broken inversion symmetry, supporting the nonreciprocity required by an intrinsic SDE. Perhaps the only exception is the kagome superconductor $\mathrm{CsV}_3\mathrm{Sb}_5$ \autocite{Le2024}, which is considered centrosymmetric at high temperatures, yet the confirmed emergence of a charge density wave (CDW) at lower temperatures breaks its inversion symmetry. Even so, the controversy surrounding its chiral order under exceedingly weak magnetic fields or strain \autocite{Guo2022a, Guo2024} hinders these kagome metals from being the ideal platform to study the interplay between spontaneous symmetry breaking and SDE. Most importantly, never before has any material system hosted both field-odd and field-free SDEs, presumably due to their different symmetry classes.

Emerging as a stellar member of the transition metal carbide (TMC) family, molybdenum carbide ($\mathrm{Mo}_2\mathrm{C}$) combines the advantages of being air-stable, mechanically strong, and easy to synthesize via chemical vapor deposition (CVD) \autocite{Wang2016, Turker2020, Zhang_2020, Ge2021}. It has gained significant attention in condensed matter physics thanks to the discovery of high-quality 2D superconductivity \autocite{Xu2015}. The material primarily exists in two phases: the orthorhombic $\alpha$-phase and the hexagonal $\beta$-phase, with recent research showing that $\beta$-$\mathrm{Mo}_2\mathrm{C}$ often exhibits higher superconducting $T_C$ and a lower residual resistance ratio (RRR) \autocite{Fan2020}. Novel theoretical proposals, including topological superconductivity and nonlinear Hall effect, have been conceived based on $\mathrm{Mo}_2\mathrm{C}$ polymorphs and their heterostructures \autocite{Shang2024, Zhao2020a, Zhang_2020, Zhao2023a}. With space groups $Pbcn$ and $P6_3/mmc$, both orthorhombic $\alpha$-phase and hexagonal $\beta$-phase have been largely considered to be centrosymmetric \autocite{Liu2015, Naher2022}. However, a recent experiment has reported an unexpected anisotropy and transitional nonreciprocity in mixed-phased $\mathrm{Mo}_2\mathrm{C}$ \autocite{Guo2025}, raising the possibility of more exotic phases of matter in this promising TMC.

In this letter, we report the discovery of SDE in CVD-grown $\mathrm{Mo}_2\mathrm{C}$ nanoflakes. By performing quantum transport measurements on lithographic microstrips, we demonstrate that SDE can occur in orthogonal crystal orientations. The field-odd SDE occurs in two samples when an in-plane magnetic field is applied perpendicular to the current flow. The diode polarity flips upon reversing the magnetic field direction, and the diode efficiency can be tuned via the field strength, reaching over 40\% at a temperature of $4$ K. The field-free SDE occurs in another sample, where the diode polarity persists under zero-field coolings as well as field coolings. In both cases, we observe no switching of SDE induced by an out-of-plane magnetic field, confirming the intrinsic nature of the phenomena. We discuss two possible mechanisms for the unexpected symmetry breakings, including domain-boundary supercurrents and CDW-like orders. While CVD-grown $\mathrm{Mo}_2\mathrm{C}$ is often regarded as a conventional $s$-wave superconductor, our findings establish it as the newest member of nonreciprocal superconductors and provide compelling reasons to consider other pairing symmetries and chiral supercurrents. With prototypical lattices, $\mathrm{Mo}_2\mathrm{C}$ polymorphs open a new avenue for the search of SDE in nominally centrosymmetric materials and shed light on our understanding of unconventional superconductors with complex microscopic structures.

\section{Results and discussion}

The $\mathrm{Mo}_2\mathrm{C}$ nanoflakes are grown via CVD using methane as the carbon source and Cu foils atop a Mo foil as the substrate at 1100℃. We characterize the thickness and flatness of the transferred flakes using atomic force microscopy (AFM) and select flat nanoflakes under $100$ nm in thickness. We fabricate Ohmic contacts with photolithography followed by Cr/Au deposition and etch the microstrip geometries via reactive ion etching (RIE). We have tried switching the order of electrode deposition and etching and observe no qualitative difference.

Orthorhombic $\alpha$-phase and hexagonal $\beta$-phase have different lattice structures while both possess inversion symmetry (Fig. 1a). In the $\beta$-phase, carbon and molybdenum atoms form distinct layered structures, and carbon atoms are distributed more disorderly compared to the $\alpha$-phase \autocite{1986Thermal}. We used atomic-scale high-angle annular dark-field scanning transmission electron microscopy (HAADF-STEM) to study the atomic structure of the CVD-grown $\mathrm{Mo}_2\mathrm{C}$ crystal (Fig. 1b). The corresponding fast Fourier transform (FFT) is demonstrated in the inset, where the pair of red circles highlight the superstructure indexed as the (020) crystal planes of $\alpha$-$\mathrm{Mo}_2\mathrm{C}$, which is consistent with previous literature result \autocite{Guo2025}. Then, we use the orthorhombic twin spots to produce the inverse FFT image, which renders a real-space map of how the $\alpha$-phase is distributed (Fig. 1c). The bright area corresponds to $\alpha$-phase, while the rest (false-colored in yellow) corresponds to $\beta$-$\mathrm{Mo}_2\mathrm{C}$. The volume ratio between $\alpha$-phase and $\beta$-phase approximates to 45:55, and the domain sizes are on the order of nanometers which are far below our mesoscopic device scale. This phase-mixing, according to the previously mentioned work \autocite{Guo2025}, could lead to spatial symmetry breaking and contribute to the phenomena we report here.

\begin{figure*}[htbp]
\centering
\includegraphics[width=0.9\textwidth]{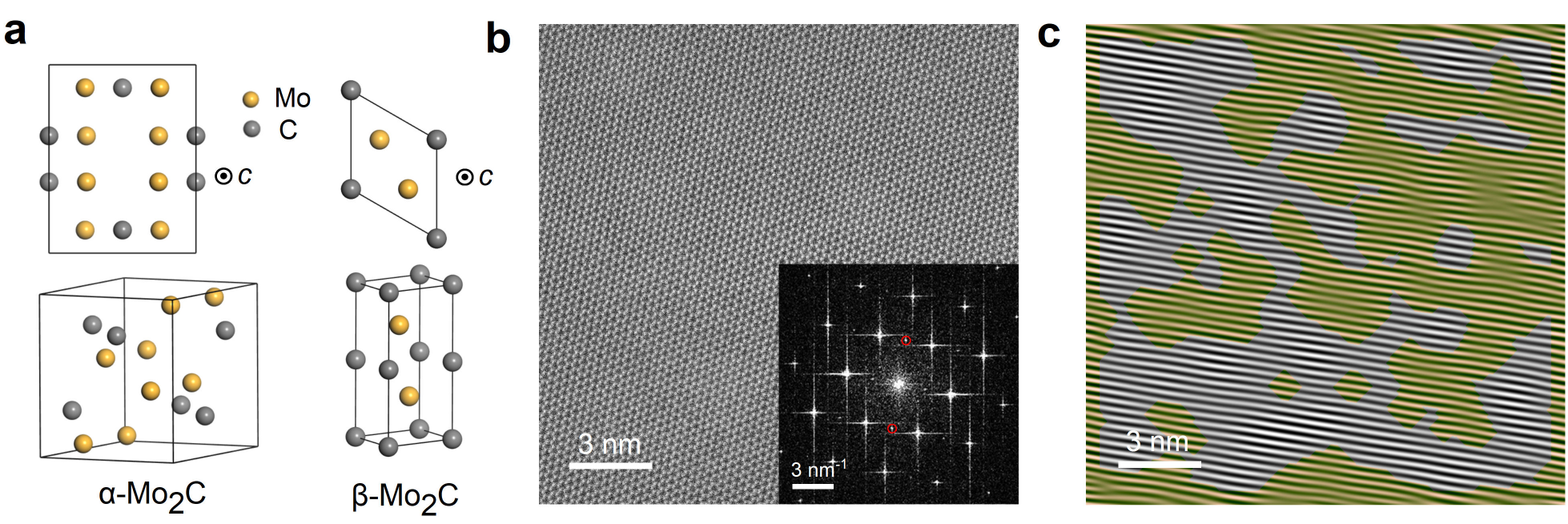}  
\caption{
\textbf{Phase mixing of $\alpha$- and $\beta$-$\mathrm{Mo}_2\mathrm{C}$.}
(a) Lattice structures of the orthorhombic $\alpha$-phase and the hexagonal $\beta$-phase $\mathrm{Mo}_2\mathrm{C}$. (b) Atomic-scale HAADF-STEM image of a $\mathrm{Mo}_2\mathrm{C}$ crystal. Inset shows the diffraction pattern after fast Fourier transform (FFT), in which the red circles highlight the orthorhombic crystal planes of $\alpha$-$\mathrm{Mo}_2\mathrm{C}$. (c) False-colored spatial map using inverse FFT displays high and low $\alpha$-phase regions in grey and yellow.}
\label{fig_1}
\end{figure*}

Figure 2a shows the optical image of Device S1 which includes microstrips of two orientations: two segments named S1-a and S1-b parallel to a natural edge of the sample and one named S1-c perpendicular to it. The thickness of the flake is measured via AFM to be 71 nm. The temperature dependence of Device S1-c shows metallic behavior down to the $10$ K range and a superconducting transition temperature of $7.9$ K (Fig. 2b), verifying our sample quality. The RRR, defined as the ratio between the sample resistance at $300$ T and the sample resistance right above the superconducting transition temperature ($10$ K in this case), is about $1.2$, which appears closer to the $\beta$ phase \autocite{Fan2020}. The square resistance is around $2.74 \; \Omega$, which places it closer to the $\alpha$-phase characteristics. The superconducting transition of Device S1-c in various in-plane magnetic field angles is shown in Fig. 2c.

\begin{figure*}[htbp]
\centering
\includegraphics[width=0.9\textwidth]{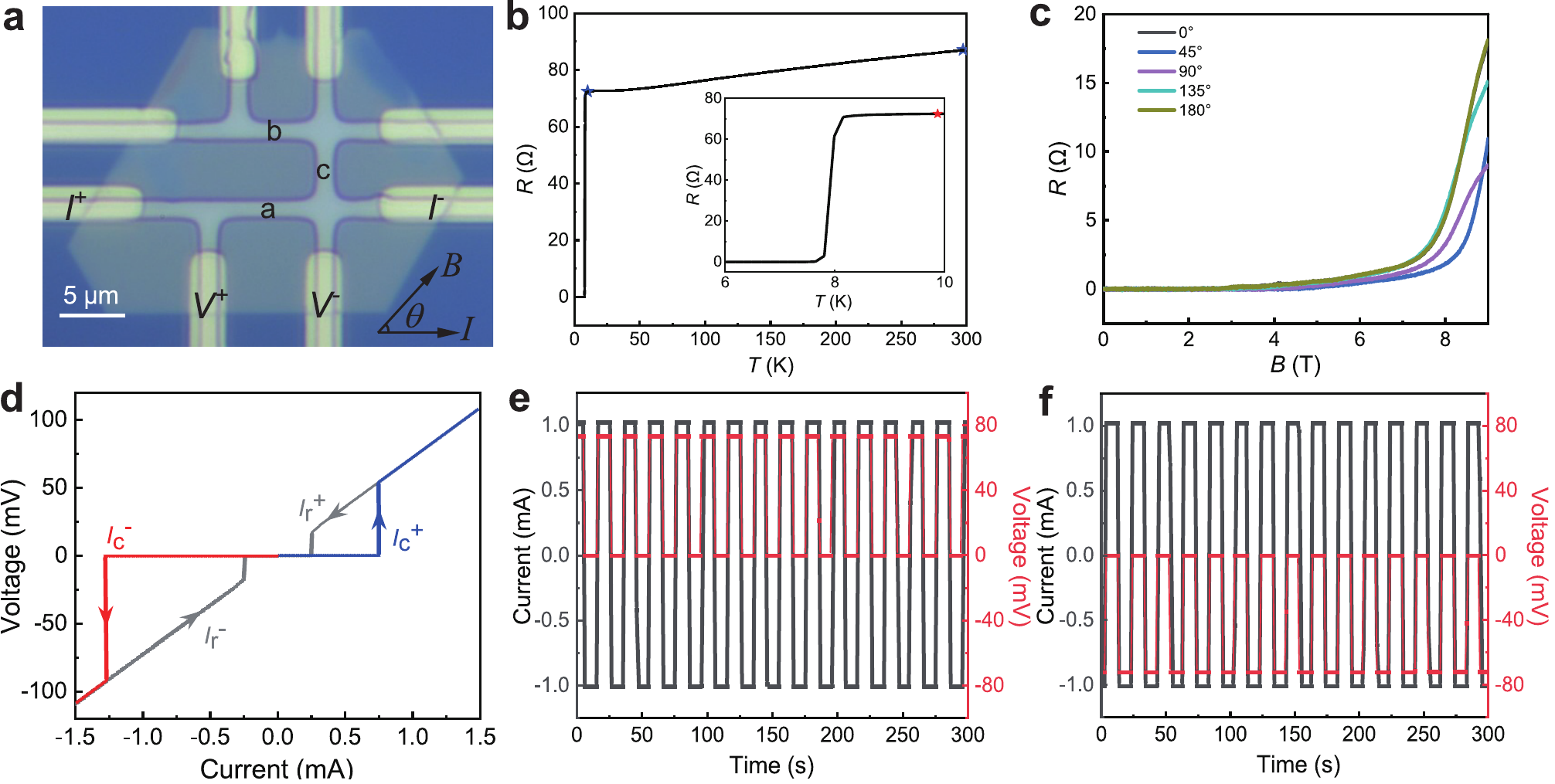}  
\caption{
\textbf{Superconducting diode effect in phase-mixed $\mathrm{Mo}_2\mathrm{C}$.}
(a) Optical micrograph and measurement schematics of Device S1. (b) Temperature dependence of electrical resistance. Inset shows a superconducting transition around $7.9$ K. (c) Superconducting transition of Device S1-c in in-plane magnetic fields, with angle defined in (a). (d) $I$-$V$ curve obtained from the configuration in (a) with $0.5$ T of in-plane field perpendicular to the microstrip exhibits hysteretic behavior, where the switching currents sweeping from zero bias to finite biases $I_{\mathrm{C}}^+$ (blue) and $I_{\mathrm{C}}^-$ (red) differ. (e, f) Switching cycles of the superconducting diode effect at $1.6$ K with opposite polarities under (e) $0.5$ T and (f) $-0.5$ T. The black curve shows the square-wave drive current (left $y$-axis) while the red curve shows the corresponding voltage drop (right $y$-axis).}
\label{fig_2}
\end{figure*}

To measure the superconducting critical current, we apply the current strictly between two ends of one sample microstrip while monitoring the voltage across the same segment using two separate electrodes, thus implementing a four-terminal measurement. We measure each segment in isolation to avoid crosstalk between two connected segments of the sample (S1-a and S1-c, for example), which could result in the inaccurate measurement of their critical currents (Supporting Information Section 1).

Figure 1d displays a representative $I$-$V$ measurement exhibiting SDE, obtained from Device S1-a with $0.5$ T of in-plane magnetic field applied perpendicular to the microstrip direction. The respective switching and retrapping currents ($I_{\mathrm{C}}^+$, $I_{\mathrm{r}}^+$) are labeled on the graph, while the sweeping directions are marked by arrows. We compare only the switching currents and not the retrapping currents for the SDE criterion to eliminate hysteretic factors such as heating. Throughout our discussion, $I_\mathrm{C}^+$ and $I_\mathrm{C}^-$ refer to the switching currents in the blue line and the red line respectively. The diode efficiency, calculated by
\begin{equation}
  \eta = \frac{I_{\mathrm{C}}^+ - |I_{\mathrm{C}}^-|}{I_{\mathrm{C}}^+ + |I_{\mathrm{C}}^-|},  \label{eqn:eta}
\end{equation} 
can reach up to $48$\% at a temperature of $3$ K. Figures 2e and f show on-off cycles of the diode operation with opposite polarities, with the applied current (black line) oscillating between $1$ mA and $-1$ mA on a square-wave signal. The measured voltage (red line) switches between zero and a finite value, displaying repeatable rectifying behavior.

To investigate the role of the magnetic field in the SDE, we vary the in-plane field keeping its direction perpendicular to the microstrip. Figure 3a shows the field-dependence of both $I_\mathrm{C}^+$ and $I_\mathrm{C}^-$ in Device S1-a. At zero field, no SDE appears, and the critical current is typically reciprocal in both flow directions. However, as we increase the field strength, $I_\mathrm{C}^+$ and $I_\mathrm{C}^-$ both decrease while their difference quickly enhances and remains pronounced in a wide range of field values before diminishing around $1$ T in both field directions. The polarity of this nonreciprocity, namely which flow direction possesses the higher critical current, switches with the sign of the magnetic field, exhibiting a field-odd SDE (Fig. 3a).

\begin{figure*}[htbp]
\centering
\includegraphics[width=0.9\textwidth]{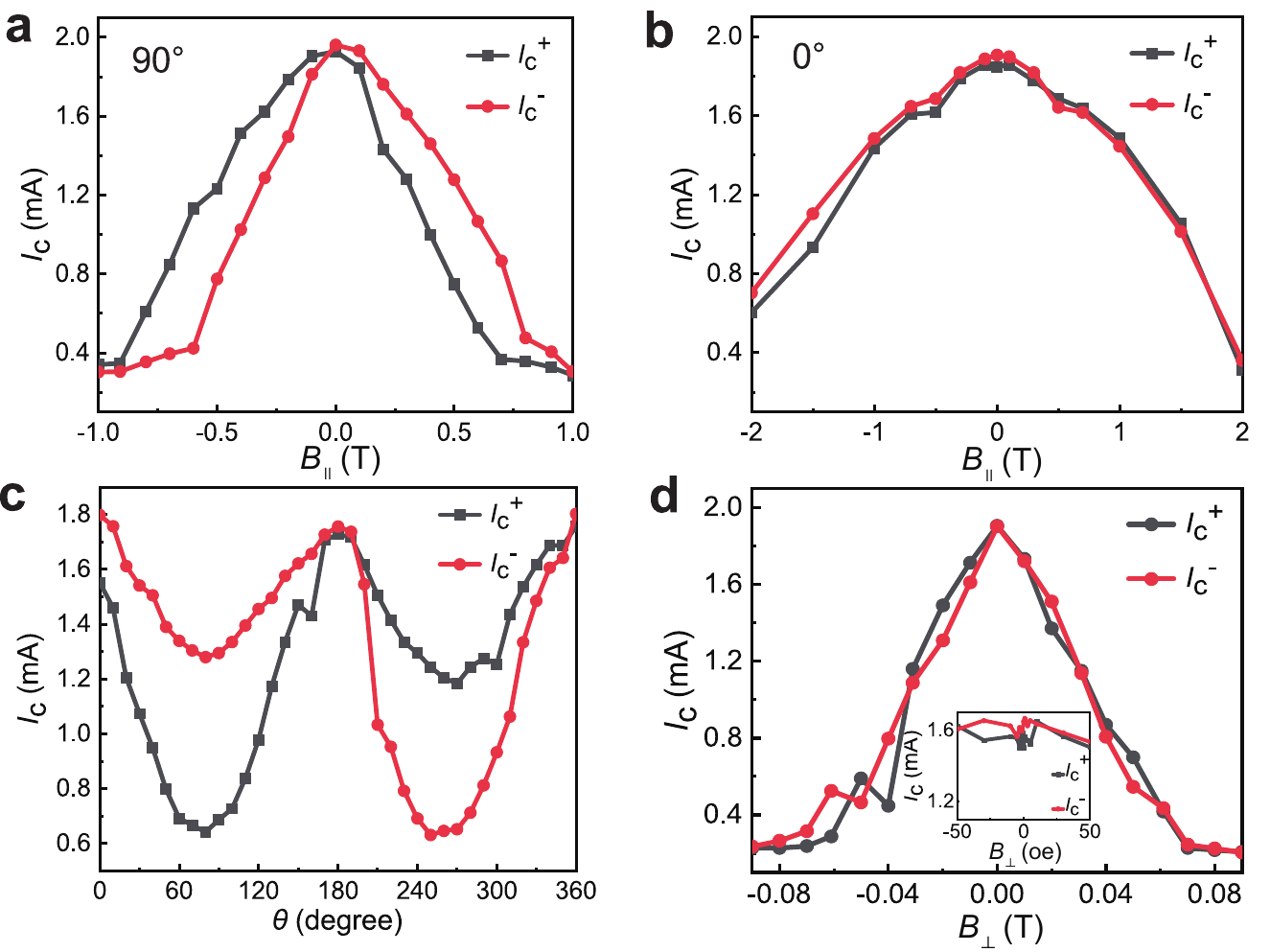}  
\caption{
\textbf{Field-odd SDE and its magnetic-field dependence.}
(a, b) In-plane field dependence of $I_\mathrm{C}^+$ and $I_\mathrm{C}^-$ with the field pointing (a) perpendicular and (b) parallel to the current direction. (c) $I_\mathrm{C}^+$ and $I_\mathrm{C}^-$ as a function of the in-plane field angle with the field magnitude fixed at $0.5$ T. (d) $I_\mathrm{C}^+$ and $I_\mathrm{C}^-$ in an out-of-plane magnetic field. Inset: small-range scan in an out-of-plane magnetic field.}
\label{fig_3}
\end{figure*}

On the other hand, when we apply an in-plane magnetic field parallel to the microstrip (direction of the current flow), we observe no significant SDE, and the critical currents stay symmetric throughout the range of field applied (Fig. 3b). To study the full angle dependence of the SDE on the field direction, we rotate our sample in-plane while keeping the magnetic field fixed at $0.5$ T, where the diode efficiency is maximized in the perpendicular in-plane field (Fig. 3c). Both $I_\mathrm{C}^+$ and $I_\mathrm{C}^-$ display an approximate two-fold in-plane anisotropy, with maxima occuring when the field is aligned with the current flow direction ($0\degree$ and $180\degree$) and minima occuring when the field is orthogonal (near $90\degree$ and $270\degree$). Some deviation from the perfect angles may be due to hysteresis in the mechanical rotation in combination with the finite width of the microstrip. At the $I_\mathrm{C}$ maxima angles, the nonreciprocity disappears as $I_\mathrm{C}^+$ (black) and $I_\mathrm{C}^-$ (red) coincide, in accordance to our parallel-field data (Fig. 3b). In the range of angles between $0\degree$ and $180\degree$, $I_\mathrm{C}^+$ stays below $I_\mathrm{C}^-$ with the highest nonreciprocity near $90\degree$; in the opposite range of angles between $180\degree$ and $360\degree$, $I_\mathrm{C}^+$ stays above $I_\mathrm{C}^-$, reversing the diode polarity, corroborating our observation in the field-sweep (Fig. 3a).

To rule out the possible scenario where SDE can appear due to edge asymmetry and vortex trapping in conventional superconducting films \autocite{Hou2023}, we apply an out-of-plane magnetic field and confirm the absence of any SDE switching. We first demonstrate on a large field scale---comparable to that of the in-plane field experiment---the lack of a consistent SDE developing (Fig. 3d). Next, we focus on the mT-scale range near zero field (Fig. 3d inset) and confirm there is no field-odd SDE there, either. The latter experimental evidence proves that the observed SDE with parallel field direction does not originate from an out-of-plane component, which usually results from imperfect sample-field alignment. Together, these sanity checks point to the SDE being intrinsic to the material.

\begin{figure*}[htbp]
\centering
\includegraphics[width=0.9\textwidth]{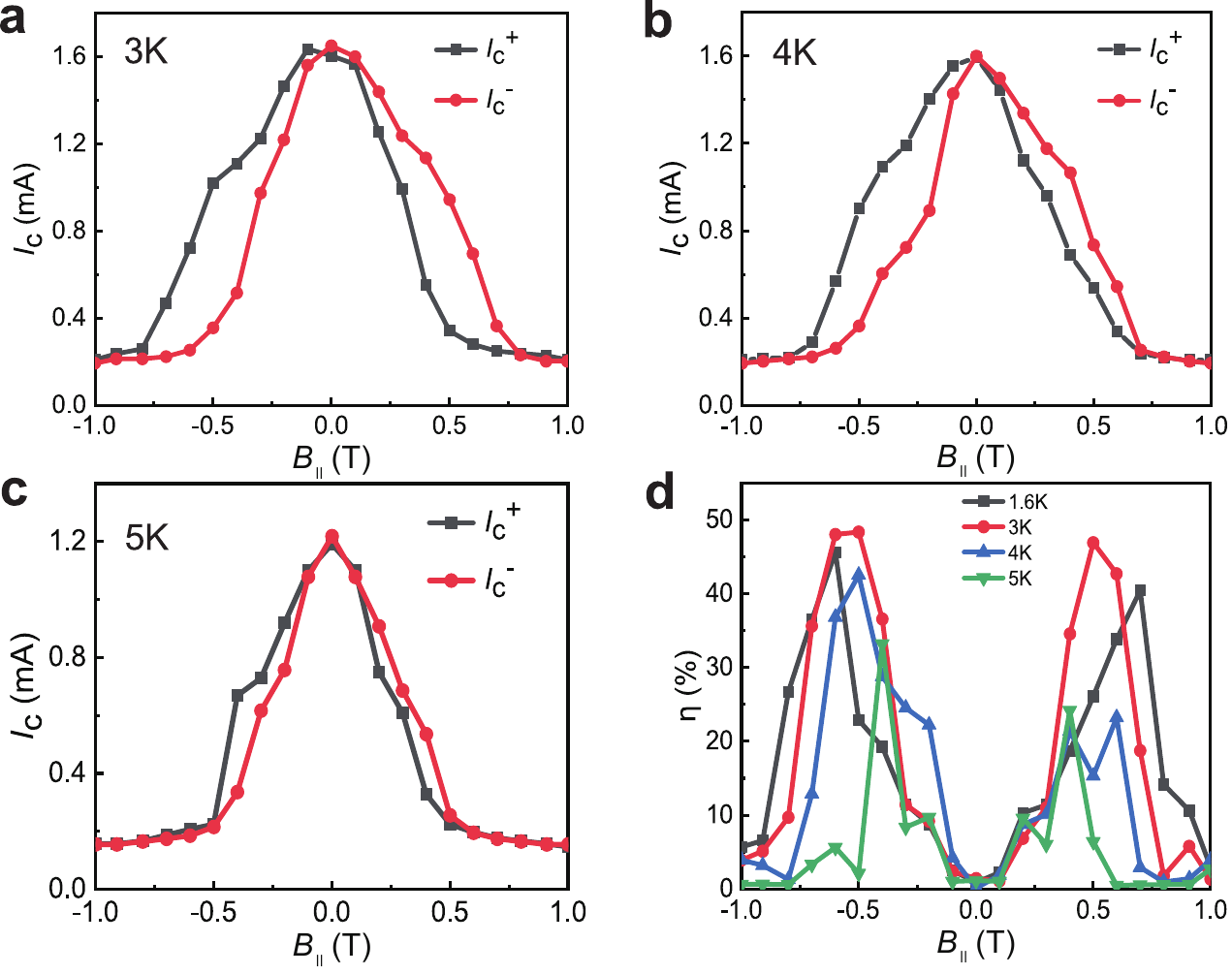}  
\caption{
\textbf{Temperature dependence of the field-odd SDE.}
(a-c) In-plane field dependence of $I_\mathrm{C}^+$ and $I_\mathrm{C}^-$ with the field perpendicular to the current at various temperatures. (d) Temperature dependence of the diode efficiency as a function of the in-plane magnetic field.}
\label{fig_4}
\end{figure*}

So far, all the datasets presented in Fig. 3 are obtained at the base temperature of our cryostat, which is around $1.6$ K. To probe the robustness of this SDE, we repeat the field-sweep in the perpendicular configuration at various temperatures. Figures 4a-c present the same measurement as Fig. 3a performed at $3$ K, $4$ K, and $5$ K, respectively. Although both $I_\mathrm{C}^+$ and $I_\mathrm{C}^-$ decrease with rising temperature, the field-odd nonreciprocity persists up to $4$ K without a significant temperature dependence (Fig. 4a and b). At $5$ K, the SDE starts to dwindle with the decreased critical currents but remains discernible with its field-odd characteristics (Fig. 4c). This wide range of diode-operating temperatures exemplifies the device's low sensitivity to thermal fluctuations, a crucial advantage for building reliable logic gates in scalable superconducting circuits. Figure 4d analyzes the diode efficiency as a function of both magnetic field and temperature and emphasizes the diode's wide working ranges in the parameter space. To verify that the observed SDE is reproducible, we have included data collected from a separate device S2 in Supporting Information Section 2.

Having established the field-odd SDE, which suggests the phase-mixed $\mathrm{Mo}_2\mathrm{C}$ breaks inversion symmetry, let us now report an even more exotic case of symmetry breaking in this system. In a third device S3 with a $T_\mathrm{C}$ of $5.9$ K, an RRR of $1.15$, and a square resistance of $10.48 \; \Omega$ (Fig. 5a and b), we observe a field-even SDE, where a sizable diode effect emerges at zero field upon cooling in the absence of any applied magnetic field (zero-field cooling). At a temperature of $1.6$ K, the diode efficiency can reach $15$\% at zero field. When an in-plane magnetic field is applied perpendicular to the current flow, the diode efficiency decreases with increased field in either direction and, again, disappears around $1$ T of field magnitude. Figures 5c and d show this behavior for two orthogonal segments of one nanoflake. Again, to rule out the ubiquitous SDE reported in Ref. \autocite{Hou2023}, we scrutinize the SDE in a small range of out-of-plane magnetic field and see the polarity of the diode persists for both crystalline directions (Fig. 5e and f). This confirms that trivial SDE related to edge asymmetry and vortex movement is not a main contributor to the field-free SDE we observe.

\begin{figure*}[htbp]
\centering
\includegraphics[width=0.9\textwidth]{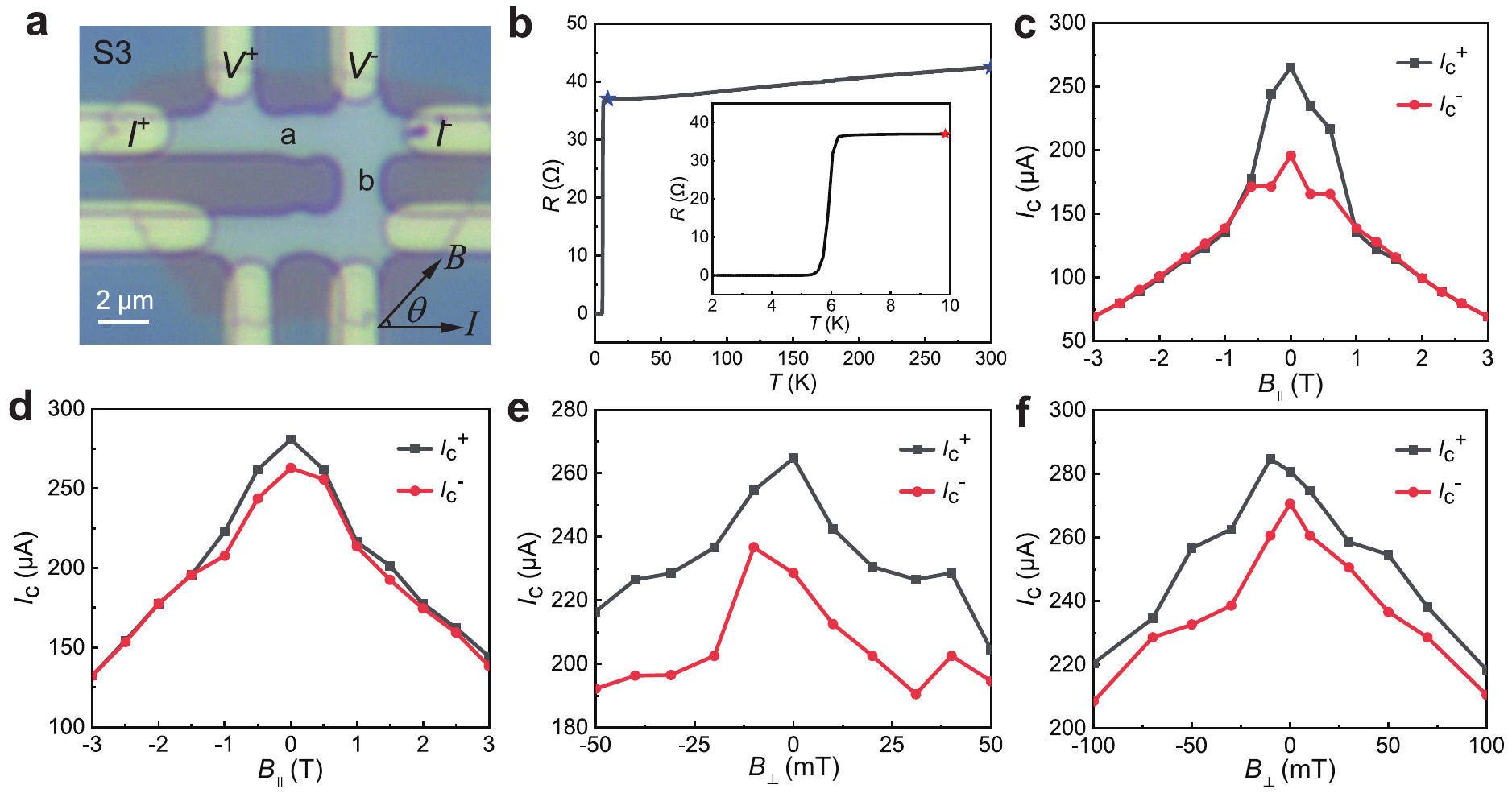}  
\caption{
\textbf{Observation of the field-free SDE in Device S3.}
(a) Optical micrograph and measurement schematics. (b) Temperature dependence of electrical resistance. Inset shows a superconducting transition around $5.9$ K. (c,d) Field dependence of $I_\mathrm{C}^+$ and $I_\mathrm{C}^-$ with the in-plane magnetic field perpendicular to the current in two orthogonal segments (c) S3-a and (d) S3-b of a nanoflake. (e,f) Small-range out-of-plane field dependence of $I_\mathrm{C}^+$ and $I_\mathrm{C}^-$ for (e) S3-a and (f) S3-b.}
\label{fig_5}
\end{figure*}

The emergence of a field-free SDE implies spontaneous TRS breaking alongside inversion symmetry breaking in the phase-mixed $\mathrm{Mo}_2\mathrm{C}$ nanoflakes. The temperature dependence of $I_\mathrm{C}^+$ and $I_\mathrm{C}^-$ displays the field-free diode effect persisting again up to $4$ K (Fig. 6a), placing the energy scale responsible for these nonreciprocities on the same order of magnitude as the superconducting gap. We notice in the field-free SDE case that the switching current can exhibit stochasticity to some extent. Figure 6b shows measurement of $I_\mathrm{C}^+$ and $I_\mathrm{C}^-$ from eight consecutive sweeps upon a separate zero-field cooling from Fig. 5c, emphasizing the robustness of the SDE through thermal cycles even though both critical currents can fluctuate in value.

To examine whether the TRS breaking originates from vortex effects such as trapped flux, we perform field-cooling with $\pm 1$ T of out-of-plane magnetic field and zero the field once cooled. Figures 6c and d show repeated measurements of the zero-field switching currents $I_\mathrm{C}^+$ and $I_\mathrm{C}^-$. All datasets for Figure 6 are collected from Device S3-b, with Device S3-a showing similar behaviors. The persistent polarity of the zero-field SDE rules out trapped flux from accidental field cooling as the root of the field-free SDE. In addition, we perform field coolings with in-plane fields (Fig. 6c and d) as well as field sweeps at base temperature (Supporting Information Fig. S8), where again no SDE polarity switch is observed despite some quantitative modification in diode efficiency. To investigate if any ferromagnetic parent phase contributes to the TRS breaking, we drive the sample above its superconducting transition via a bias current and perform magnetic field sweeps. The resulting magnetoresistance displays no apparent hysteresis, confirming $\mathrm{Mo}_2\mathrm{C}$ to be non-magnetic in the normal regime and therefore pinning the TRS-breaking mechanism inside the superconducting phase itself (Supporting Information Fig. S6).

\begin{figure*}[htbp]
\centering
\includegraphics[width=0.9\textwidth]{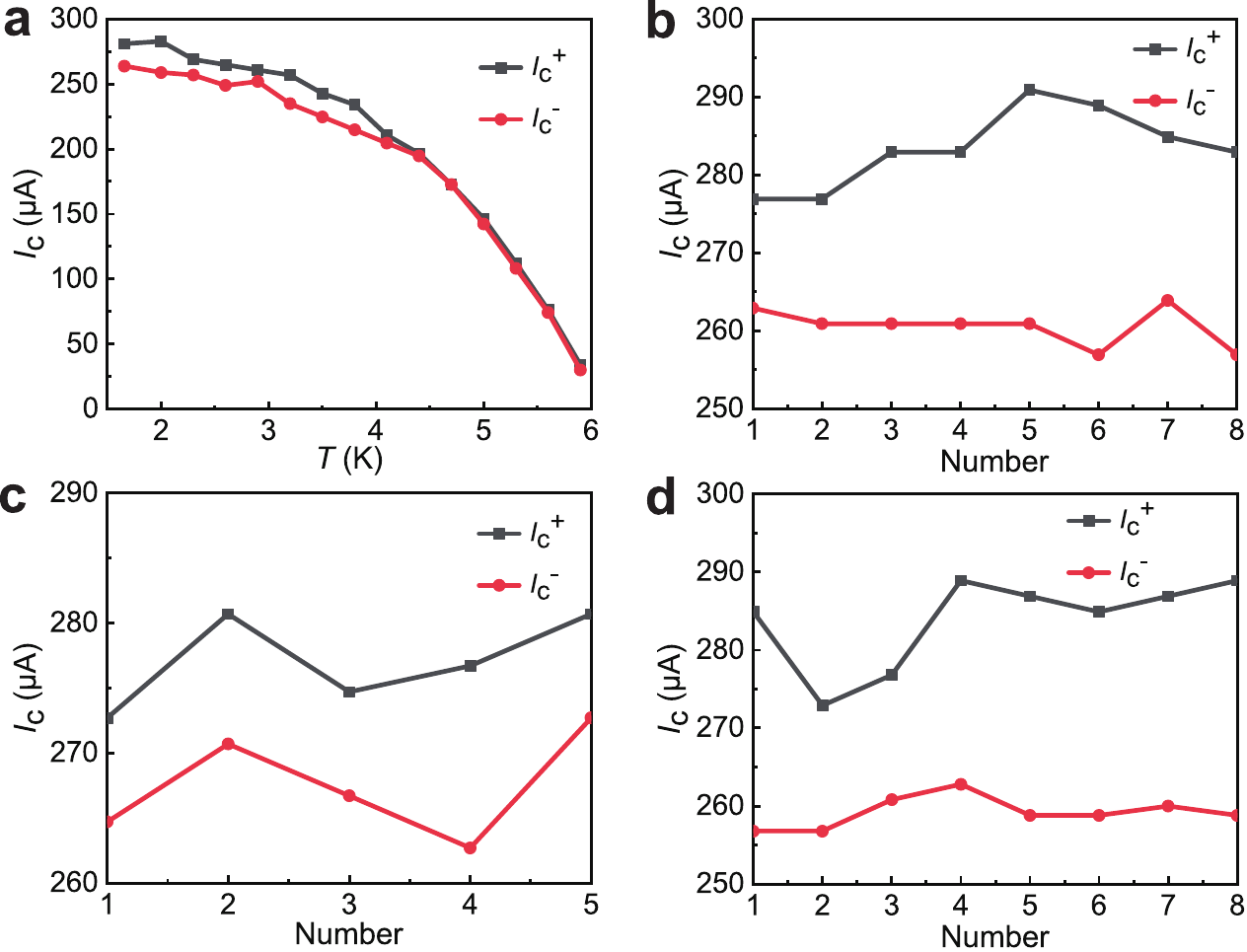}  
\caption{
\textbf{Temperature and field-cooling dependence of the field-free SDE.}
(a) Temperature dependence of $I_\mathrm{C}^+$ and $I_\mathrm{C}^-$ below the superconducting transition. (b-c) Repeated measurements of $I_\mathrm{C}^+$ and $I_\mathrm{C}^-$ upon (b) zero-field cooling, (c) $-1$ T out-of-plane field cooling, and (d) $1$ T out-of-plane field cooling.}
\label{fig_6}
\end{figure*}

It is important to note that the field-odd and field-free SDEs arise in separate $\mathrm{Mo}_2\mathrm{C}$ nanoflakes, but both behaviors emerge from the same growth process. This leads us to the central scientific surprise in our findings: the origin of the broken symmetries in superconducting $\mathrm{Mo}_2\mathrm{C}$. The emergence of SDE requires a combination of inversion symmetry breaking and TRS breaking \autocite{Nadeem2023}. In our field-odd superconducting diodes, the in-plane magnetic field breaks the TRS. However, both $\alpha$- and $\beta$-phase $\mathrm{Mo}_2\mathrm{C}$ possess centrosymmetric lattice structures where inversion symmetry is intact \autocite{Liu2015, Naher2022}. A recent study sheds light on this intrigue by showing that $\mathrm{Mo}_2\mathrm{C}$ could contain polymorph interation that breaks such centrosymmetry and results in a two-fold in-plane anisotropy in its magnetoresistance as well as nonreciprocal transport during the superconducting transition \autocite{Guo2025}. We perform similar measurements and reproduce the two-fold in-plane anisotropy as well as the asymmetric second-harmonic response (Supporting Information Section 3). STM data in \autocite{Guo2025} further corroborates this spatial symmetry breaking with a periodic modulation of electron density of states, hinting that the TRS could be simultaneously broken via this emerging electron order. Such a many-body correlation could certainly contribute to our observed broken symmetries, especially if it supports chiral orders. However, a spontaneous TRS-breaking mechanism usually renders its chirality coercible by an external magnetic field, especially during field coolings. The unchanging polarity of our field-free SDE requires a mechanism more engraved in the supercurrent flow throughout the phase-mixed sample.

The recent discovery in kagome superconductor $\mathrm{CsV}_3\mathrm{Sb}_5$ reveals an interesting scenario where, driven by the strong electron-electron interaction, chiral superconducting domains give rise to boundary supercurrents, leading to a global zero-field SDE \autocite{Le2024}. The diode efficiency in $\mathrm{CsV}_3\mathrm{Sb}_5$ is under $10$\% and the polarity switches upon thermal cycling, suggesting a weak intrinsic chiral order in the sample. This could be understood if the different domains originate from the same superconducting order parameter with mere microscopic variations. In our phase-mixed $\mathrm{Mo}_2\mathrm{C}$, however, we have a mixture of two distinct superconducting phases with different superfluid densities, different critical temperatures, and potentially different pairing symmetries \autocite{Shang2024, Zhao2020a}. This creates a more drastic contrast at their phase boundaries and a richer network of symmetry-broken edge states than kagome metals.

That said, one also cannot rule out the possibility that phase-mixed $\mathrm{Mo}_2\mathrm{C}$ could host a CDW order similar to $\mathrm{CsV}_3\mathrm{Sb}_5$, which coexists with domain-boundary supercurrents. Such periodic modulations of electron density can shift the center of charge away from the center of the lattice, possibly demolishing the inversion center. Previous work has demonstrated the competition between CDW and superconductivity in Cr-doped $\mathrm{Mo}_2\mathrm{C}$ \autocite{Li2021b}. If, indeed, CDW is the origin of these SDEs, then the intertwined charge density and superconducting orders could lead to exotic phenomena such as pair density waves (PDW). This opens an exciting future direction for low-temperature scanning tunnelling microscopy \autocite{Gu2023, Zhao2023b, Han2025a, Bi2024}, which are also well-suited to study the microscopic superconducting domains. Besides, scanning nano-SQUID magnetometry could be a powerful tool to visualize the boundary supercurrents \autocite{Finkler2010, Zhu2025a, Iguchi2023}. In terms of transport experiments, fabricating interference rings to study the Little-Partks effect through different parts of a $\mathrm{Mo}_2\mathrm{C}$ nanoflake would shed light on the interaction between local symmetry breakings and the global SDE \autocite{Liao2022, Zhang2024d, Almoalem2024, Wan2024}. Finally, regarding material synthesis, a natural direction would be to purify the phases through meticulous control of growth conditions so that we can examine the behaviors of individual $\mathrm{Mo}_2\mathrm{C}$ phases.

In conclusion, our discoveries of field-odd and field-free SDEs in CVD-grown $\mathrm{Mo}_2\mathrm{C}$ nanoflakes challenge the traditional understanding of symmetry in nominally centrosymmetric materials. Through multi-directional transport measurements, we observe field-tunable superconducting diode efficiencies exceeding $40$\% at $4$ K. In another sample, we witness robust zero-field SDE with persistent polarity under zero-field coolings as well as field coolings. By scanning in an out-of-plane magnetic field, we exclude mundane effects such as vortex pinning from asymmetric sample edges. These field-odd and field-free non-reciprocal responses point toward novel symmetry-breaking mechanisms; they raise questions about the interaction between local domain interfaces and global symmetry classes on a mesoscopic scale, which is at the heart of modern devices near the quantum limit. From an applicational perspective, we establish $\mathrm{Mo}_2\mathrm{C}$ as a versatile platform for air-stable superconducting diode operations at liquid-helium temperature. Most importantly, this work not only expands the family of non-reciprocal superconductors but also provides a new paradigm for uncovering hidden asymmetries in crystal materials, paving the way for the next generation of unconventional superconducting devices.

\printbibliography

\section{Experimental}

\textbf{CVD Growth of $\mathrm{Mo}_2\mathrm{C}$ nanoflakes:} 3 Cu foils ($99.8$\% purity, 25 µm thick) are cut into $10 \times 10$ $\mathrm{mm}^2$ pieces and placed on top of a Mo foil ($99.95$\% purity, 50 µm thick) of the same size. Subsequently, the substrates are heated above 1100 °C in a horizontal tube furnace under 100 sccm $\mathrm{H}_2$ and 100 sccm Ar. 3.5 sccm $\mathrm{CH}_4$ is then introduced into the chamber at ambient pressure to initiate the growth of flakes. After 35 minutes, $\mathrm{CH}_4$ supply and heating are terminated, and the sample is allowed to cool naturally to room temperature under a $\mathrm{H}_2$ and Ar environment.

\textbf{Transfer of nanoflakes:} A thin layer of poly-(methyl methacrylate) (PMMA, weight-averaged molecular mass Mw = 600,000, 4 wt\% in ethyl lactate) is first spin-coated on the surface of $\mathrm{Mo}_2\mathrm{C}$ crystals at 3,000 rpm. for 1 min and cured at 180 °C for 2 min; the PMMA-coated samples are immersed into a 1M ${(\mathrm{NH}_4)}_2\mathrm{S}_2\mathrm{O}_8$ aqueous solution for etching the underlying Mo and Cu substrates. Then the PMMA-coated $\mathrm{Mo}_2\mathrm{C}$ crystals are stamped onto target substrates, such as $\mathrm{SiO}_2$/Si and TEM grids, and warm acetone (55 °C) is used to dissolve the PMMA layer and obtain clean $\mathrm{Mo}_2\mathrm{C}$ flakes.

\textbf{Device fabrication:} After transferring the nanoflakes onto Si substrates with 290-nm-thick $\mathrm{SiO}_2$, we etch the nanoflakes into microstrip devices using RIE with Ar and $\mathrm{SF}_6$. Contact electrodes are patterned on the top surface by direct-write photolithography followed by thermal evaporation of Cr/Au (5/80 nm).

\textbf{Transport measurements:} We perform the majority of the measurement in a Kelvince variable temperature insert (VTI) with up to $9$ T of magnetic field. The VTI provides a temperature range from $1.6$ K to $300$ K and is equipped with an electronically controlled rotating sample stage. A small portion of the measurement is also conducted in a Quantum Design Physical Property Measurement System (PPMS) offering similar capabilities. The DC current bias is sourced from a Keithley 2460 source meter. We use a NF 5645 lock-in amplifier to source the AC current excitation as well as measure the AC voltage. 

\section{Author Contributions}
H. Ren conceived the experiment. W. Gao, M. Li, and Q. Zhang conducted the material synthesis under D. Geng's supervision. W. Gao fabricated the devices under the supervision of H. Ren, K. Fan, and F. Yang. W. Gao and P. Zhu performed the STEM measurement. W. Gao, K. Fan, and J. Cheng performed the transport measurements under the supervision of H. Ren, S. Ding, W. Hu, and F. Yang. W. Gao and H. Ren analyzed the data and prepared the manuscript with input from all authors.

\section*{Acknowledgements}

We thank the Fundamental and Interdisciplinary Disciplines Breakthrough Plan of the Ministry of Education of China (JYB2025XDXM410). This work is broadly supported by the National Natural Science Foundation of China; S. Ding acknowledges grant no. 52373250, and D. Geng acknowledges grant no. 524B2011. We thank Gang Chen from Peking University for a helpful theoretical discussion. We also thank Yu Pan and Yufeng Huang from Tianjin University for their perspectives on the STEM result.

\section*{Supporting information}

The Supporting Information is available: Measurement of critical currents from adjacent microstrips, field-odd SDE repeated in a separate sample, in-plane magneto-anisotropy and second-harmonic measurement, and investigating the origin of broken TRS in the field-free SDE (PDF).

\end{spacing}

\end{document}